\documentclass[12pt]{article}
\usepackage[utf8]{inputenc}
\usepackage[T1]{fontenc}
\usepackage{mathptmx}              
\usepackage[numbers,square]{natbib}
\usepackage{amsmath, amssymb, amsfonts}
\usepackage{bm}
\usepackage{graphicx}
\usepackage{booktabs}
\usepackage{url}
\usepackage[margin=1in]{geometry}
\usepackage[colorlinks=true,linkcolor=blue,citecolor=blue,urlcolor=blue]{hyperref}

\newcommand{\pkg}[1]{\textsf{#1}}
\newcommand{\code}[1]{\texttt{#1}}
\newcommand{\doi}[1]{\href{https://doi.org/#1}{\texttt{doi:#1}}}

\title{\pkg{smoothbp}: Fast Bayesian Hierarchical Piecewise Regression with Smoothed Transitions and Spike-and-Slab Model Selection}

\author{
  Aidan Bindoff \\[4pt]
  Wicking Dementia Research \& Education Centre, \\
  College of Health \& Medicine, \\
  University of Tasmania \\[4pt]
  \texttt{aidan.bindoff@utas.edu.au}
}

\date{}

\begin{document}

\maketitle

\begin{abstract}
Piecewise regression models are essential for identifying structural changes in longitudinal or spatial data across diverse scientific domains. While standard approaches often assume sharp, instantaneous transitions and single, non-hierarchical breakpoints, many real-world phenomena exhibit gradual, smoothed transitions that vary systematically across groups. We introduce \pkg{smoothbp}, an R package for fast, Bayesian hierarchical piecewise regression featuring logistic-smoothed transitions. By implementing a bespoke Metropolis-within-Gibbs sampler in Rust, \pkg{smoothbp} combines exact conjugate updates for linear terms with Hamiltonian Monte Carlo (HMC) transitions for non-linear location and sharpness parameters. \pkg{smoothbp} natively supports multiple change-points, random intercepts, random change-point timing, and structural covariates on all segment parameters. It also incorporates Kuo and Mallick (1998) spike-and-slab priors for automatic inference on the number of active breakpoints via the \code{smoothbp\_ss} function. We document the sampler, validate parameter recovery and calibration through simulation-based calibration and interval-coverage studies, and contrast \pkg{smoothbp} against the existing software landscape across R, Python, Julia, and MATLAB, demonstrating its competitive efficiency against general-purpose probabilistic programming languages like \pkg{brms} and specialized packages like \pkg{mcp}.
\end{abstract}


\bigskip

\section{Introduction}

Identifying structural shifts or breakpoints in data over time or space is a ubiquitous challenge across scientific domains. In public health, interrupted time series and regression discontinuity designs are used to evaluate policy interventions. In ecology, abrupt shifts in climate proxies indicate regime changes. In financial econometrics, structural breaks signify market events. Piecewise (or segmented) regression models directly estimate the locations of these shifts and the resulting changes in trajectory.

However, three significant modeling challenges frequently arise in modern applications:
\begin{enumerate}
    \item \textbf{Hierarchical Structure}: Data are rarely independent; they are often collected in nested or grouped structures (e.g., repeated measures within patients). The underlying baseline, as well as the exact timing of a structural shift, may vary hierarchically across these groups.
    \item \textbf{Transition Smoothness}: Traditional piecewise regression imposes a hard, instantaneous ``kink'' at the breakpoint. Many natural transitions occur gradually over an identifiable window.
    \item \textbf{Unknown Number of Breaks}: While exploratory approaches test fixed numbers of breaks, robust Bayesian inference seeks to jointly quantify the probability of the number of shifts alongside their parameters.
\end{enumerate}

To address these challenges, we introduce \pkg{smoothbp}, an R package that implements Bayesian hierarchical piecewise regression utilizing logistic-smoothed transitions. The approach is adapted from the smooth-transition piecewise regression of \citet{baconwatts1971}, who joined two intersecting straight lines with a smooth hyperbolic-tangent transition and estimated it by Bayesian analysis. \pkg{smoothbp} adopts the same smooth-transition principle with a logistic parameterisation (an affine reparameterisation of the hyperbolic tangent) and generalises it to multiple change-points, hierarchical change-point timing, and covariates on every segment parameter. Powered by a bespoke Rust backend, \pkg{smoothbp} pairs exact conjugate Gibbs updates with Hamiltonian Monte Carlo (HMC) to deliver inference for multiple breakpoints, covariates on all parameters, and built-in Kuo \& Mallick spike-and-slab regularization for model selection.

This paper describes \pkg{smoothbp} version~0.2.7, the current release on the Comprehensive R Archive Network (CRAN).

\section{The Software Landscape}

The challenge of estimating unknown breakpoints has spurred numerous software implementations across modern computational environments. To contextualize the contribution of \pkg{smoothbp}, we briefly review the current software landscape across R, Python, Julia, and MATLAB. (Full search methodology provided in Appendix A).

\subsection{R Environment}
R possesses the richest ecosystem for piecewise regression.
\begin{itemize}
    \item \pkg{segmented} \citep{muggeo2003}: The standard frequentist package. It supports random effects via \code{segmented.lme}, allowing for random intercepts and random breakpoint locations. However, transitions are strictly instantaneous, and selecting the number of breakpoints requires iterative information criterion checks.
    \item \pkg{mcp} \citep{lindelov2020}: A powerful Bayesian package wrapping JAGS. \pkg{mcp} supports multiple change-points, various likelihood families, and random change-point timing. It assumes hard, instantaneous transitions, and while it computes LOO/WAIC, it lacks native spike-and-slab sparsity.
    \item \pkg{brms} \citep{burkner2017}: General-purpose probabilistic programming via Stan. While capable of fitting smoothed, hierarchical transitions via custom non-linear formulas (\code{nl = TRUE}), \pkg{brms} requires the user to manually derive and specify the logistic-smoothed geometry, bounded priors, and meticulous initialization to prevent HMC chains from becoming trapped in unidentifiable boundary regions.
\end{itemize}

\subsection{Python, Julia, and MATLAB}
Outside of R, hierarchical segmented modeling typically requires custom implementation.
\begin{itemize}
    \item \textbf{Python}: Packages like \pkg{piecewise-regression} offer straightforward non-hierarchical piecewise fitting. Signal processing libraries like \pkg{ruptures} excel at detecting shifts but do not perform hierarchical regression. For true mixed-effects, users must rely on custom probabilistic models built in \pkg{PyMC} or \pkg{NumPyro}.
    \item \textbf{Julia}: \pkg{MixedModels.jl} provides state-of-the-art frequentist mixed-effects fitting, but users must manually update the segmented design matrix via optimization wrappers. We are not aware of an automated Bayesian multi-breakpoint tool akin to \pkg{mcp} or \pkg{smoothbp} (search current to mid-2025; see Appendix A).
    \item \textbf{MATLAB}: The Statistics and Machine Learning Toolbox provides \code{fitlme}, but simultaneous estimation of random effects and an unknown, non-linear breakpoint relies on custom non-linear optimization routines (e.g., \code{fmincon}).
\end{itemize}

No existing package provides a turnkey, formula-driven interface that combines all three features (hierarchical change-point timing, logistic smoothing, and spike-and-slab breakpoint selection) without requiring the user to hand-code the model geometry. \pkg{smoothbp} fills this niche while keeping wall-clock time competitive with general-purpose Hamiltonian frameworks.

\section{Statistical Framework}

\subsection{Notation and the Smoothed Piecewise Model}

We use $i$ to index observations, $j \in \{1, \dots, J\}$ to index groups (e.g.\ subjects), and $k \in \{1, \dots, K\}$ to index change-points. Observation $i$ belongs to group $j[i]$ and is taken at time $\tau_i$. The logistic function is written $\operatorname{logit}^{-1}(x) = (1 + e^{-x})^{-1}$.

A model with $K$ change-points is

\begin{equation}
y_{i} = b_0 + u_{0,j[i]} + b_1\,(\tau_{i} - \omega_{1,j[i]}) + \sum_{k=1}^K \delta_{k}\,(\tau_{i} - \omega_{k,j[i]})\,\operatorname{logit}^{-1}\!\left( \rho_k\,(\tau_{i} - \omega_{k,j[i]}) \right) + \epsilon_{i},
\end{equation}

where $b_0$ is a reference level, $u_{0,j} \sim \mathcal{N}(0, \sigma_u^2)$ is an optional random intercept, $b_1$ is the baseline slope, $\delta_k$ is the change in slope at change-point $k$, $\omega_{k,j}$ is its location, $\rho_k$ is its transition sharpness, and $\epsilon_{i} \sim \mathcal{N}(0, \sigma^2)$.

Two features of this parameterisation are important and were chosen to match the implementation rather than textbook convention.

\paragraph{Centering of the baseline slope.} The baseline slope term is $b_1(\tau_i - \omega_{1,j})$, that is, time is measured \emph{relative to the first change-point}, not relative to the origin. A direct consequence is that $b_0$ is \emph{not} the intercept at $\tau = 0$; it is the conditional mean of the response at the first change-point, $\tau = \omega_{1,j}$, where every $\delta_k$ term vanishes by construction (each shift term is centred on its own $\omega_k$). This makes $b_0$ directly interpretable as the pre-shift level and stabilises the sampler, because $b_0$ and the $\delta_k$ no longer trade off through a remote origin. When $K = 0$ the model degenerates to ordinary linear regression and the baseline slope reverts to $b_1\tau_i$.

\paragraph{Smoothness.} As $\rho_k \to \infty$ the logistic transition converges to a Heaviside step, recovering the classical hard-kink segmented model; finite $\rho_k$ models a gradual transition of identifiable width. Estimating $\rho_k$ rather than fixing it lets the data express how abrupt each shift is.

\subsection{Hierarchical Timing, Parameterisation, and Identifiability}

\pkg{smoothbp} supports random effects on the change-point location. For group $j$, the timing of event $k$ is distributed around the population mean:
\begin{equation}
\omega_{k,j} \sim \mathcal{N}(\omega_k, \sigma_{\mathrm{re},k}^2).
\end{equation}
This accommodates settings where a shared intervention or environmental trigger affects all groups but biological or logistical delays induce heterogeneous onset times.

\paragraph{Parameterisation.} By default the random change-points are sampled in a centred parameterisation. When the group-level variance $\sigma_{\mathrm{re},k}$ is poorly identified or small relative to the data, the joint $(\omega_{k,j}, \sigma_{\mathrm{re},k})$ geometry can develop a funnel that degrades HMC mixing. For such cases \pkg{smoothbp} exposes \code{reparameterise = "omega"}, a fully non-centred parameterisation of the change-point random effects~\cite{papaspiliopoulos2007}, which is the recommended remedy when funnel pathology is diagnosed.

\paragraph{Identifiability and ordering.} With more than one change-point, the likelihood is invariant to relabelling: swapping $(\omega_1, \delta_1, \rho_1)$ and $(\omega_2, \delta_2, \rho_2)$ leaves the mean function unchanged, so an unconstrained posterior can be multimodal with the change-point labels exchanged. This is the standard label-switching problem of mixture and change-point models. \pkg{smoothbp} addresses it by encouraging the user to impose non-overlapping, ordered priors on the $\omega_k$. The helper \code{space\_omega\_priors()} constructs a set of truncated-normal priors with disjoint support over a user-specified time window, and per-parameter \code{lb}/\code{ub} bounds can be set directly. We recommend ordered, non-overlapping $\omega$ priors whenever $K > 1$. A related pitfall, where a change-point drifts beyond the observed time range ($\omega > \max_i \tau_i$) into an unidentifiable region, is likewise avoided by bounding $\omega$ within the data window.

\subsection{Spike-and-Slab Selection via Kuo and Mallick (1998)}

Identifying the correct number of breakpoints is reframed as a variable selection problem. In \code{smoothbp\_ss}, we apply the Kuo \& Mallick spike-and-slab prior \citep{kuomallick1998} to each slope change $\delta_k$, decomposing it into a binary inclusion indicator $\gamma_k$ and a continuous slab coefficient $\beta_{k}$:

\begin{align}
\delta_k &= \gamma_k \cdot \beta_{k} \\
\gamma_k &\sim \text{Bernoulli}(\pi) \\
\beta_{k} &\sim \mathcal{N}(0, \sigma_{\text{slab}}^2)
\end{align}

If $\gamma_k = 0$, the $k$-th structural shift is zeroed out (the ``spike''). If $\gamma_k = 1$, the slope change is estimated from the diffuse ``slab''. The posterior mean of $\gamma_k$ is the Posterior Inclusion Probability (PIP). A hyperprior $\pi \sim \text{Beta}(a, b)$ can be specified to allow the data to inform global sparsity. The \code{pip()} accessor returns a data frame with one row per candidate shift and a column of posterior inclusion probabilities; column labels follow the design-matrix convention (e.g.\ \code{gamma\_delta1\_(Intercept)} is the PIP for the intercept term of the first slope change), so that PIPs for covariate-modulated shifts are individually addressable.

\subsection{Model Selection and Averaging via Inclusion Probabilities}
\label{sec:pip}

The spike-and-slab output supports two complementary uses, and \pkg{smoothbp} furnishes both from a single fit.

The first is explicit model selection. Selecting the change-points whose PIP is at least $0.5$ yields the \emph{median probability model} (MPM). This threshold is not an arbitrary heuristic: \citet{barbieriberger2004} showed that the MPM is optimal for prediction under squared-error loss in orthogonal and nested designs, a result later extended to broader correlated-design settings by \citet{barbieri2021}. In \pkg{smoothbp} the MPM is read directly off \code{pip(fit\_ss)} as the set of shifts with $\text{PIP} \ge 0.5$. One caveat carries over from that literature: when candidate terms are strongly correlated, several can sit just above $0.5$ and the MPM may over-select; for well-separated change-points the $\delta_k$ are close to orthogonal, so this is rarely an issue in practice, but for closely-spaced candidates it is worth inspecting the joint inclusion pattern rather than the marginal PIPs alone.

The second use is Bayesian model averaging (BMA). Because the inclusion indicators $\gamma_k$ are sampled jointly with the continuous parameters, every posterior draw corresponds to a particular subset of active change-points, so any quantity computed from the draws --- in particular the posterior predictive mean and its credible intervals --- is automatically a model average in which each candidate model is weighted by its posterior probability. A user who forms predictions from the full set of \pkg{smoothbp} draws is therefore not committing to a single selected model but reporting the BMA prediction, which propagates uncertainty about \emph{which} change-points are real into the forecast. The MPM and the BMA prediction answer different questions --- ``which single model should I report?'' versus ``what is the best prediction, accounting for model uncertainty?'' --- and both are available without refitting.

\paragraph{A caution on indicator mixing.} The reliability of any PIP rests on the inclusion indicators $\gamma_k$ being well mixed. In version~0.2.7 the indicators are updated by the Kuo \& Mallick scheme conditional on the current slab coefficient, and we have found that under the diffuse default slab this update can mix slowly: an excluded coefficient drawn from a wide slab fits poorly, so the indicator is reluctant to re-enter, and the chain can dwell in one inclusion state for long stretches. In practice the PIPs still discriminate active from inactive shifts, but their numerical values should be read as indicative rankings rather than precisely calibrated probabilities unless the per-indicator effective sample size is checked and found adequate. Users can improve mixing by running longer chains, tightening the slab toward the expected effect size, or comparing models by leave-one-out cross-validation or marginal likelihood (Section~6) as a cross-check. A collapsed (Rao-Blackwellised) indicator update, which integrates the slab coefficient out analytically and is expected to mix substantially better, is planned for a future release; we report the current behaviour honestly rather than defer the feature.

\subsection{Default Priors}

The default priors are weakly informative on an order-unity scale; their second argument is a standard deviation throughout. They are: $\delta_k \sim \mathcal{N}(0, 1.5)$ on slope changes, a half-normal $\mathcal{N}(0, 1)$ truncated to the data window for $\omega$, and $\rho \sim \mathcal{N}(3, 2)$ truncated to $\rho > 0$ for sharpness. The variance components ($\sigma$, $\sigma_u$, and $\sigma_{\mathrm{re}}$) take inverse-gamma priors, a family the conjugate Gibbs update requires; the default is $\mathrm{InvGamma}(1, 1)$. Because the $\delta$, $\omega$, and $\rho$ priors are stated on the natural scale of the data, they are weakly informative only when the response and time axis are of order unity; on data of very different scale the user should rescale or adjust them, since \pkg{smoothbp} does not standardise inputs automatically. The sharpness prior deserves comment: $\rho$ is the parameter the data most often fail to pin down, because a transition that is sharp relative to the observation spacing is indistinguishable from one that is sharper still. The default places its mass on moderate, identifiable transition widths while permitting near-instantaneous kinks; for intervention designs where the transition is known to be abrupt, users can fix $\rho$ with \code{fixed()} (Section~5) or tighten its prior. We recommend a brief prior predictive check of the implied transition width for any new application.

\section{Computational Details}

General-purpose NUTS samplers (like Stan) must evaluate the gradient of the entire joint log-posterior with respect to all parameters. In piecewise models, local correlations between the baseline level, random effects, and slopes can severely constrain the step size.

\pkg{smoothbp} uses a specialized Metropolis-within-Gibbs sampler (more precisely, HMC-within-Gibbs, since the non-linear blocks use Hamiltonian updates) implemented in safe Rust.
\begin{itemize}
    \item \textbf{Conjugate Linear Updates}: Conditioned on the non-linear parameters ($\omega$, $\rho$) and the variance components, the linear parameters ($b_0$, $u_{0,j}$, $b_1$, $\beta_{k}$) form a Gaussian system. \pkg{smoothbp} jointly updates the population level and all group-level random intercepts $u_{0,j}$ in a single block using exact Gibbs sampling. Blocking these correlated quantities together avoids the slow-mixing ridge between the population intercept and the group-level deviations that element-wise updates would suffer. This is distinct from Neal's funnel: the centred $(u, \sigma_u)$ geometry can still be challenging when $\sigma_u$ is very small, which is why the non-centred option of Section~3.2 exists for the change-point random effects.
    \item \textbf{HMC Non-Linear Updates}: The location $\omega$ and sharpness $\rho$ lack conjugate full conditionals and are updated block-wise via Hamiltonian Monte Carlo, using forward-mode automatic differentiation implemented natively in Rust for exact leapfrog gradients. The \code{test-gradients.R} suite checks these analytic gradients against finite differences.
    \item \textbf{Per-subject Adaptation for Random Change-points}: When change-point timing is hierarchical, each group's location parameter is adapted independently. This per-subject adaptation is essential for good mixing when the subject-level change-point distributions differ, and a joint Gibbs translation step couples the population and subject-level location parameters to keep their sum well mixed (Section~\ref{sec:calibration}).
    \item \textbf{Per-block Divergence Reporting}: The fit object exposes \code{n\_divergent\_by\_block}, a list giving the number of divergent HMC transitions attributable to the subject-level, $\omega$, and $\rho$ blocks separately. This makes it possible to localise a convergence problem to the parameter block responsible rather than reporting a single opaque count, and is used directly in the validation of Section~\ref{sec:calibration}.
    \item \textbf{Fixed Parameter Bypasses}: Parameters known \emph{a priori} (e.g., intervention times) can be pinned with the \code{fixed()} helper, instructing the sampler to bypass gradient calculations and proposals for them entirely.
\end{itemize}

The default HMC target acceptance probability is \code{target\_accept = 0.9}.

\section{Using smoothbp}

\pkg{smoothbp} requires a Rust toolchain (Cargo and \code{rustc >= 1.65.0}) at install time; the package vendors its Rust dependencies so no network access is needed during compilation, but a first install takes roughly 10--15 minutes to build the backend. This system requirement, declared in \code{SystemRequirements}, distinguishes \pkg{smoothbp} from pure-R packages and is worth communicating to users before they run \code{install.packages()}.

The API mirrors standard R formula syntax via \code{lme4}-style conventions.

\begin{verbatim}
# Fit a model with 3 candidate breakpoints and spike-and-slab selection
library(smoothbp)

fit_ss <- smoothbp_ss(
  formula = response ~ time,
  b0      = ~ 1 + (1 | subject),
  b1      = ~ 1,
  deltas  = list(~ 1, ~ 1, ~ 1),
  omega   = list(~ 1, ~ 1, ~ 1),
  rho     = list(~ 1, ~ 1, ~ 1),
  data    = my_data,
  spike   = prior_spike_slab(pi = 0.2, learn_pi = TRUE),
  priors  = smoothbp_priors(
    omega = space_omega_priors(K = 3, tau_min = 0, tau_max = 10)
  )
)

# Examine Posterior Inclusion Probabilities
pip(fit_ss)
\end{verbatim}

For intervention analyses like Regression Discontinuity Designs or Stepped-Wedge trials, users can exploit the \code{fixed()} wrapper to pin locations and sharpness:

\begin{verbatim}
omega = list(fixed(3.0)),
rho   = list(fixed(100)) # Approximate a hard kink
\end{verbatim}

A fitted model can be re-run with modified settings via \code{update()}; sampler tuning (\code{step\_om}, \code{step\_rho}, \code{target\_accept}) and the random seed are stored in the fit object and reused unless explicitly overridden, so carefully tuned runs are not silently reset.

\subsection{Convenience Functions}

A few helpers streamline the most common workflows. \code{space\_omega\_priors(K, tau\_min, tau\_max)} returns a ready-to-use list of $K$ ordered, non-overlapping truncated-normal priors for the change-point locations: the prior means are spaced evenly across the time window (with the edges padded inward) and each is bounded to the window, which is the recommended way to impose the ordering that multi-change-point models require to avoid label switching. \code{fixed(value)} pins a parameter at a known value and bypasses its sampling entirely --- the mechanism behind the intervention-analysis idiom shown above, where a fixed $\omega$ encodes a known intervention time and a large fixed $\rho$ approximates a hard kink. \code{tab\_smoothbp()} collects posterior summaries from one or more fitted models into a single publication-ready table with parameters on rows and models in columns (rendered through \pkg{gt}, falling back to \code{knitr::kable()}), in the spirit of \code{sjPlot::tab\_model()}. Finally, \code{simulate\_smoothbp()} generates data from the model and stores the data-generating values as a \code{"true\_params"} attribute, so that \code{true\_params()} and \code{recovery\_plot()} can overlay the posterior against the known truth --- the loop behind the package's parameter-recovery tests and a convenient template for users designing their own simulation studies.

\subsection{Convergence and Calibration for Random Change-points}
\label{sec:calibration}

Random change-point timing is the most demanding feature of the model, so we characterise its behaviour directly rather than asserting good performance. All results in this section come from reproducible scripts: \code{tools/part3\_validation.R} for the mixing diagnostics, and \code{tools/sbc\_ecdf\_bands.R} with \code{tools/sbc\_recovery\_diagnostic.R} for the calibration and recovery studies.

\paragraph{Mixing.} On a representative random change-point problem (low-prior-fraction regime: $J = 20$ balanced groups, modest between-group variance $\sigma_{\mathrm{re}}$, four chains), the sampler mixes efficiently across the hierarchical timing parameters (Table~\ref{tab:mixing}). The subject-level locations $\omega_j$ reach a mean bulk effective sample size (ESS) of 1768 with $\widehat{R} = 1.00$, the population change-point $\omega$ an ESS of 256, and the between-group standard deviation $\sigma_{\mathrm{re}}$ an ESS of 2970, with no divergent transitions in the subject or $\omega$ blocks. The population change-point is the slowest-mixing of these (an order of magnitude below $\sigma_{\mathrm{re}}$), because it is strongly correlated with the baseline level; an ESS of 256 from 16{,}000 draws is ample for the posterior mean but is the parameter to watch when reporting extreme tail quantiles of $\omega$.

\begin{table}[t]
\centering
\caption{Random change-point mixing, low-prior-fraction regime ($J = 20$, balanced design, modest $\sigma_{\mathrm{re}}$), four chains of 4000 post-warmup draws. ``Subject $\omega_j$'' is the mean across the 20 group-level locations.}
\label{tab:mixing}
\begin{tabular}{lrr}
\toprule
Parameter & Bulk ESS & $\widehat{R}$ \\
\midrule
Subject $\omega_j$ (mean) & 1768 & 1.00 \\
Population $\omega$       &  256 & 1.01 \\
$\sigma_{\mathrm{re}}$   & 2970 & 1.00 \\
\bottomrule
\end{tabular}
\end{table}

\begin{figure}[t]
\centering
\includegraphics[width=\textwidth]{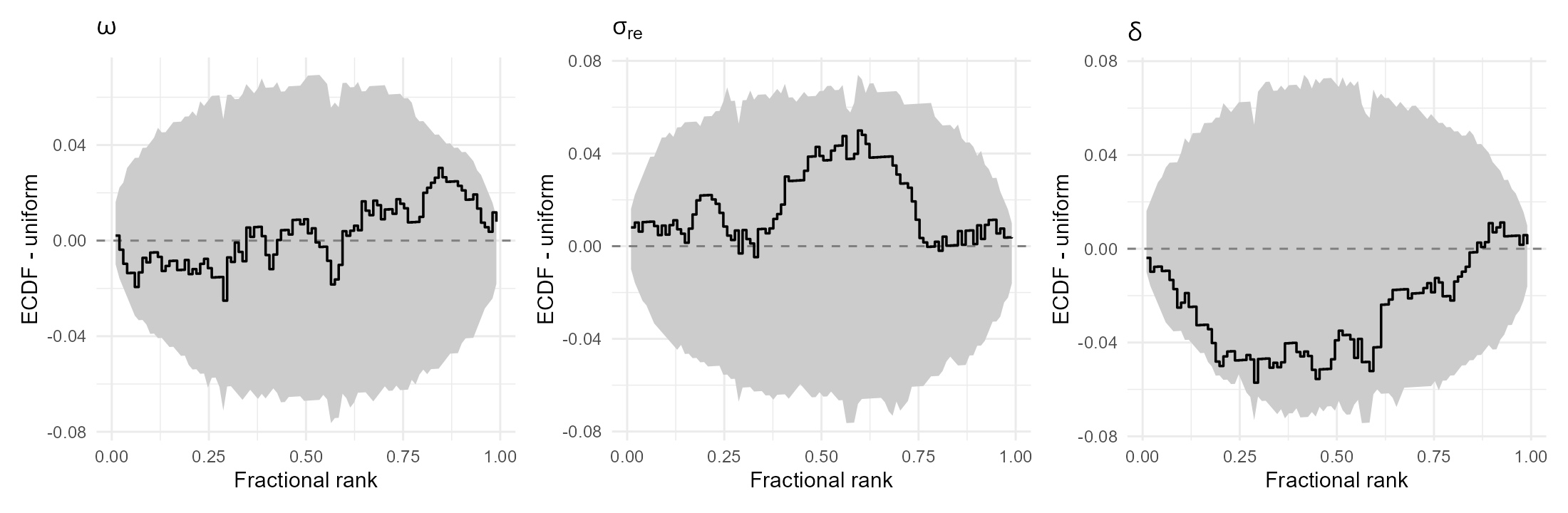}
\caption{Simulation-based calibration for the random change-point model ($\rho$ fixed, 500 replicates). Rank ECDFs minus the uniform diagonal for $\omega$, $\sigma_{\mathrm{re}}$, and $\delta$ stay within the $95\%$ simultaneous confidence bands, indicating calibrated posteriors.}
\label{fig:sbc}
\end{figure}

\paragraph{Simulation-based calibration and recovery.}
As a strong correctness check we ran simulation-based calibration (SBC) for the random change-point model under a fully valid design: every estimated parameter ($b_0$, $b_1$, $\delta$, $\omega$, $\sigma$, $\sigma_{\mathrm{re}}$) was drawn from its prior and the model was fitted with those same priors, over $500$ replicates ($J = 6$ groups of $18$ observations, two chains of $3000$ iterations, $L = 100$ thinned near-independent draws per replicate, \code{target\_accept = 0.99}). We assess rank uniformity with simultaneous empirical-CDF confidence bands \citep{sailynoja2022}, which are substantially more powerful than the binned chi-square test \citep{talts2018}. With the transition sharpness $\rho$ fixed at its generating value the sampler is divergence-free ($0/500$), and the rank ECDFs for the change-point location $\omega$, the slope change $\delta$, and the between-group standard deviation $\sigma_{\mathrm{re}}$ all stay within the $95\%$ simultaneous bands (Figure~\ref{fig:sbc}).

When $\rho$ is instead estimated, its weakly-identified geometry induces divergent transitions in coarse regimes (here $379/500$ replicates); restricting to the converged fits (no divergences, bulk ESS $\ge 200$) restores rank uniformity for $\omega$, $\delta$, and $\sigma_{\mathrm{re}}$. The miscalibration under estimated $\rho$ therefore reflects contaminated fits rather than estimator bias, and is removed by fixing $\rho$ or raising \code{target\_accept}, consistent with the sharpness limitation discussed in Section~7.

A complementary recovery study (150 divergence-free replicates) confirms the point estimates and intervals directly (Table~\ref{tab:coverage}): posterior bias is negligible for every parameter ($|\text{bias}| \le 0.02$ prior standard deviations), and empirical coverage of nominal $90\%$ intervals is at or above nominal throughout, with no anti-conservative parameter. The change-point parameters are calibrated; the intercept $b_0$ and baseline slope $b_1$ are mildly conservative (their $90\%$ intervals cover approximately $95\%$), which overstates uncertainty slightly rather than understating it.

\begin{table}[t]
\centering
\caption{Posterior recovery and interval coverage from the valid SBC (150 divergence-free replicates, $\rho$ fixed). Bias is in units of the prior standard deviation; coverage is the empirical fraction of replicates whose generating value fell inside the nominal $50\%$ and $90\%$ posterior intervals.}
\label{tab:coverage}
\begin{tabular}{lrrr}
\toprule
Parameter & Bias (prior SD) & Cov.\ 50\% & Cov.\ 90\% \\
\midrule
$b_0$ (intercept)        &  0.02 & 0.59 & 0.95 \\
$b_1$ (baseline slope)   &  0.01 & 0.57 & 0.97 \\
$\delta$ (slope change)  &  0.00 & 0.48 & 0.88 \\
$\omega$ (change-point)  & -0.01 & 0.50 & 0.91 \\
$\sigma$ (residual SD)   &  ---  & 0.50 & 0.89 \\
$\sigma_{\mathrm{re}}$   &  ---  & 0.55 & 0.89 \\
\bottomrule
\end{tabular}
\end{table}

\paragraph{Localising the sharpness divergences.} These divergences belong to the sharpness geometry, not the hierarchical-timing machinery: in a control with the change-point held \emph{fixed} (no $\omega$ random effect) the $\rho$-block divergences persist, isolating them from the random-effect structure. We therefore treat them as a known, separable limitation of the sharpness parameter (Section~7).

\section{Benchmarking}

We benchmark \pkg{smoothbp} against \pkg{brms} (Stan) and \pkg{mcp} (JAGS) in the package's \code{brms-comparison} vignette, which is fully reproducible and covers four scenarios of increasing complexity: a single change-point with a random intercept; a covariate on $\omega$; a single-group comparison against \pkg{mcp}; and a two-breakpoint single-group model. The figures below are from Scenario~1 (a random-intercept model, $J = 25$ subjects, 10 observations each, four chains of 2000 iterations run sequentially on a single core) on one machine; effective-sample-size-per-second is machine-, data-, and configuration-dependent, so we present one representative run and provide the vignette for full reproduction.

\begin{figure}[t]
\centering
\includegraphics[width=\textwidth]{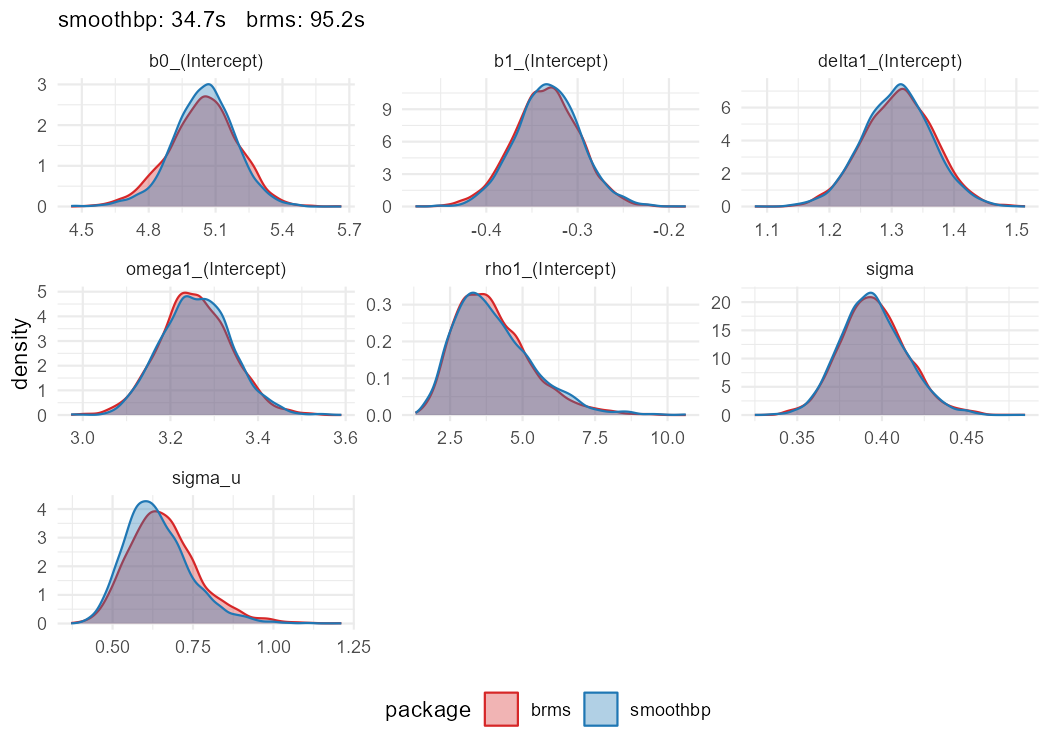}
\caption{Scenario~1 marginal posterior densities for \pkg{smoothbp} and \pkg{brms} under identical data and matched priors. The two samplers' marginals overlay closely for every parameter except $\sigma_u$; the visible separation is in the tails of the weakly-identified $\rho_1$ and, more substantially, in $\sigma_u$, whose marginal differs because the inverse-gamma and half-$t$ priors cannot be matched exactly (Table~\ref{tab:w1}).}
\label{fig:overlay}
\end{figure}

\paragraph{Posterior agreement.} On identical data and matched priors, the two samplers produce marginal posteriors that overlay closely (Figure~\ref{fig:overlay}). The structural-parameter posterior means coincide to within $0.07$ of a posterior standard deviation (Table~\ref{tab:agreement}), and the standardised 1-Wasserstein analysis below shows that the agreement extends to the full marginal shape for every parameter except the random-effect standard deviation $\sigma_u$, where the differing variance-prior families leave a residual offset that we examine in detail.

\begin{table}[t]
\centering
\caption{Scenario~1 structural-parameter posterior means (single clean run, matched priors). Agreement across the full marginals, including the variance components $\sigma$ and $\sigma_u$, is quantified in Table~\ref{tab:w1}.}
\label{tab:agreement}
\begin{tabular}{lrr}
\toprule
Parameter & \pkg{brms} mean & \pkg{smoothbp} mean \\
\midrule
$b_0$        &  5.045 &  5.048 \\
$b_1$        & -0.335 & -0.332 \\
$\delta_1$   &  1.312 &  1.309 \\
$\omega_1$   &  3.257 &  3.260 \\
$\rho_1$     &  3.972 &  4.005 \\
\bottomrule
\end{tabular}
\end{table}

\paragraph{Distributional distance.} Agreement of means, and even of standard deviations, does not by itself establish that two samplers target the same distribution. We therefore quantify the full marginal discrepancy with the 1-Wasserstein (earth-mover's) distance, standardised by the pooled posterior standard deviation so that it is reported in scale-free units. Any finite run carries Monte Carlo error, so a standardised distance of exactly zero is not expected even for identical targets; we calibrate against this floor with a within-sampler reference --- the distance between two independent \pkg{smoothbp} runs (seeds 31 and 32) at the same draw count. For a fair comparison all priors are matched, including the variance components: because \pkg{smoothbp}'s conjugate variance update requires an inverse-gamma family, both packages were given diffuse variance priors (\pkg{smoothbp} an \code{InvGamma(0.001, 0.001)}, \pkg{brms} a wide half-$t$). We avoid calling the inverse-gamma ``non-informative'': $\mathrm{InvGamma}(\epsilon, \epsilon)$ concentrates mass near zero and is in fact informative for a weakly identified variance component \citep{gelman2006}, a point that bears directly on the $\sigma_u$ comparison below.

Table~\ref{tab:w1} reports both distances. For six of the seven parameters --- including the residual standard deviation $\sigma$, whose cross-package distance ($0.037$) matches its within-sampler reference ($0.034$) --- the cross-package distance is at or near the reference, indicating marginals that are indistinguishable up to Monte Carlo error. The single exception is the random-effect standard deviation $\sigma_u$ ($0.21$ versus a $0.02$ reference). This parameter is identified only by the 25 groups, so its posterior remains partly shaped by the prior; since the inverse-gamma and half-$t$ families cannot be made identical, a small residual offset survives where the data do not dominate. That this is a prior-specification effect rather than a sampler discrepancy is established by $\sigma$ itself: identified by 250 observations, it agreed only after its prior was matched (an earlier run with mismatched default variance priors gave $W_1^{\text{cross}} = 0.46$ for $\sigma$, collapsing to $0.04$ once matched). These are marginal comparisons; combined with the shared parameterisation they are strong evidence that the two implementations target the same posterior, though they do not test the joint distribution directly.

\begin{table}[t]
\centering
\caption{Standardised 1-Wasserstein distance (in pooled posterior-SD units) between the \pkg{smoothbp} and \pkg{brms} Scenario~1 marginals ($W_1^{\text{cross}}$), and a within-sampler Monte Carlo reference from two \pkg{smoothbp} seeds ($W_1^{\text{null}}$), under matched diffuse priors. Computed by the \code{brms-comparison} vignette.}
\label{tab:w1}
\begin{tabular}{lrr}
\toprule
Parameter & $W_1^{\text{cross}}$ & $W_1^{\text{null}}$ \\
\midrule
$b_0$        & 0.086 & 0.068 \\
$b_1$        & 0.068 & 0.072 \\
$\delta_1$   & 0.050 & 0.031 \\
$\omega_1$   & 0.046 & 0.076 \\
$\rho_1$     & 0.042 & 0.053 \\
$\sigma$     & 0.037 & 0.034 \\
$\sigma_u$   & 0.210 & 0.017 \\
\bottomrule
\end{tabular}
\end{table}

\paragraph{Efficiency.} On this scenario \pkg{smoothbp} completed in 28\,s against 90\,s for \pkg{brms}. The \pkg{brms} figure includes one-time Stan model compilation, which is amortised to zero across repeated fits, so this single-run wall-clock gap overstates the steady-state difference; the fairer comparison excludes compilation, and we report sampling-only times alongside it in the vignette. The advantage is therefore real but should not be read as a fixed multiple. The per-second efficiency is parameter-dependent and reflects the hybrid sampler design (Table~\ref{tab:ess}): \pkg{smoothbp}'s conjugate block updates give it markedly higher effective sample size per second on the linear and variance parameters ($\sigma$: 119 vs 32 ESS/s; $\sigma_u$: 71 vs 4; $\delta_1$: 85 vs 24), whereas \pkg{brms}'s global NUTS is more efficient per second on the non-linear change-point $\omega_1$ (14 vs 31) and on the intercept $b_0$ (3.5 vs 5.2), which mixes slowly in \pkg{smoothbp} because it is strongly correlated with the change-point location; the two are essentially tied on the sharpness $\rho_1$ (29 vs 31).

\begin{table}[t]
\centering
\caption{Scenario~1 bulk ESS per second and wall-clock time (single clean run). Both packages ran four chains sequentially on a single core; the \pkg{brms} time includes one-time Stan model compilation.}
\label{tab:ess}
\begin{tabular}{lrr}
\toprule
Parameter & \pkg{smoothbp} ESS/s & \pkg{brms} ESS/s \\
\midrule
$b_0$        &   3.5 &  5.2 \\
$b_1$        &  27.9 & 22.5 \\
$\delta_1$   &  84.8 & 23.9 \\
$\omega_1$   &  14.3 & 30.9 \\
$\rho_1$     &  29.3 & 31.2 \\
$\sigma$     & 119.2 & 32.4 \\
$\sigma_u$   &  70.5 &  4.4 \\
\midrule
Wall-clock (s) & \textbf{28.1} & \textbf{90.0} \\
\bottomrule
\end{tabular}
\end{table}

Against \pkg{mcp}, \pkg{smoothbp} matches the random-intercept and random change-point capabilities while additionally offering smoothed transitions and the Kuo \& Mallick spike-and-slab architecture for collapsing superfluous segments, which \pkg{mcp} does not provide.

For model comparison beyond spike-and-slab PIPs, \pkg{smoothbp} integrates with the \pkg{loo} package for approximate leave-one-out cross-validation and supports marginal-likelihood estimation via \pkg{bridgesampling}, so that PIP-based selection and information-criterion or Bayes-factor comparison are both available and complementary.

\section{Discussion}

We present \pkg{smoothbp}, a tool for Bayesian hierarchical piecewise regression. By modeling smoothed logistic transitions rather than hard, instantaneous kinks, the software provides a general framework for structural change, and by pairing conjugate Gibbs updates for linear terms with bounded HMC for the non-linear terms it mixes well across the regimes we have tested, including the demanding random change-point case validated in Section~\ref{sec:calibration}.

\textbf{Interpretability}: A practical strength of the logistic-smoothed parameterisation is that its parameters map onto quantities practitioners actually want to report. Away from the transitions the mean function is asymptotically linear, so the segment slopes --- $b_1$ before the first change-point and $b_1 + \sum_k \delta_k$ after the last --- are ordinary, directly interpretable rates of change, and each $\delta_k$ is the change in slope attributable to event $k$. The location $\omega_k$ is the point of maximum change, the inflection of the $k$-th transition where the trajectory turns most sharply, and is reported on the natural scale of the time variable; it answers ``when did the change occur?'' directly. When the sharpness $\rho_k$ is itself estimable, the entire mean function is smooth and differentiable everywhere, in contrast to hard-kink models whose derivative is undefined at the breakpoint. This is not merely intuitively appealing. A globally differentiable mean is mathematically convenient --- the gradient with respect to every parameter exists in closed form --- and computationally advantageous, because it is exactly the smoothness that the Hamiltonian sampler exploits to move efficiently: the property that makes the model easy to interpret is the same one that makes it efficient to fit.

\textbf{Relationship to additive models}: Faced with grouped, non-linearly trending longitudinal data, many analysts reach instead for a generalized additive mixed model (GAMM; e.g.\ the \pkg{mgcv} package, \citealp{wood2017}), attracted by its automatic smoothness selection through penalised splines. A GAMM fits such data flexibly and chooses its effective degrees of freedom from the data, but the fitted smooth is a wiggly function whose shape must be read from a plot: it does not, by itself, estimate \emph{when} a trajectory changed or by \emph{how much} its slope shifted. \pkg{smoothbp} trades the GAMM's general flexibility for a parameterisation in which exactly those quantities --- the change-point locations $\omega_k$ and the slope changes $\delta_k$ --- are explicit parameters carrying posterior credible intervals, while still selecting the number of active change-points automatically through the spike-and-slab prior. When the scientific question concerns the timing and magnitude of structural change, this is the more natural summary; when it concerns an arbitrary smooth trend, a GAMM is.

\textbf{Limitations}: \pkg{smoothbp} is specialized for Gaussian likelihoods and logistic-smoothed transitions. It does not support autoregressive error structures (AR/ARMA), non-Gaussian exponential-family likelihoods (e.g., Poisson, Binomial), or fully non-parametric functional segments (e.g., splines); for those, general frameworks like \pkg{brms} or \pkg{mcp} remain appropriate. A further consequence of the conjugate design is that variance components are restricted to the inverse-gamma family: the exact Gibbs update that makes the linear and variance blocks fast precludes the half-Cauchy or half-$t$ priors often preferred for group-level standard deviations, and this matters most in designs with few groups, where the prior on $\sigma_u$ is not dominated by the data \citep{gelman2006}. The same parametric commitment that buys interpretability is itself a limitation relative to GAMMs: where the response is a genuinely smooth, non-monotone function not well approximated by a small number of linear segments, or where multiple additive smooth terms, tensor-product interactions, or non-Gaussian families are required, a GAMM is the more appropriate tool, and \pkg{smoothbp} is best reserved for data whose structure is plausibly piecewise-linear with a few smoothed transitions. The transition-sharpness parameter $\rho$ is weakly identified in data with coarse temporal resolution and can exhibit residual HMC divergences (Section~\ref{sec:calibration}); when the transition width is not of direct scientific interest, fixing $\rho$ or tightening its prior is the pragmatic remedy. As with all multi-change-point models, ordered priors on the change-point locations are needed for $K > 1$ to avoid label switching. Finally, the spike-and-slab selection of version~0.2.7 inherits the slow indicator mixing of the Kuo \& Mallick scheme under a diffuse slab (Section~\ref{sec:pip}): the PIPs separate active from inactive shifts but are not yet precisely calibrated at default settings, and we recommend checking indicator effective sample size and cross-checking with leave-one-out or marginal-likelihood comparison. A collapsed indicator update that is expected to resolve this is planned for a subsequent release.

\textbf{Conclusion}: For researchers requiring estimation of transition timing, hierarchical structure on change-points, and exploratory spike-and-slab screening of multiple breakpoints within a single formula interface, \pkg{smoothbp} offers a fast option in the R ecosystem whose core estimation and calibration we have validated directly, and whose selection machinery we report on candidly pending a sampler improvement.

\section*{Acknowledgements}

The software, its documentation, and this manuscript were developed with the assistance of agentic AI coding tools, used as pair programmers under the author's direction and supervision. The original package architecture (the Rust MCMC backend, the formula interface, and the first manuscript draft) was developed in collaboration with Google's Project Antigravity assistant running Gemini 3.1 Pro. Refinements to the random change-point sampler, the accompanying validation suite (\code{tools/part3\_validation.R}), the pre-submission code review, and the present manuscript revision were carried out in collaboration with Anthropic's Claude models. All code and text were independently reviewed, tested, and validated by the author, who takes full responsibility for the correctness and reproducibility of the work in accordance with COPE authorship guidelines.

\section*{Appendix A: Software Search Strategy}
To characterise the state of piecewise-regression software, a literature and repository review was conducted using Google Scholar and domain-specific package indices (CRAN, PyPI, JuliaHub). Search strings combined permutations of \textit{("hierarchical piecewise regression" OR "mixed-effects segmented regression" OR "Bayesian change point")} with \textit{(software OR package OR python OR R OR julia OR MATLAB)}. The review confirmed that while tools for single breakpoints or non-hierarchical multiple breaks exist across languages, no package offers a turnkey, formula-driven interface that simultaneously provides Bayesian multi-breakpoint estimation, hierarchical change-point timing, logistic-smoothed transitions, and built-in spike-and-slab selection without requiring the user to hand-code the model. A user of a general probabilistic-programming framework (\pkg{brms}, \pkg{PyMC}, \pkg{Turing.jl}) can of course assemble such a model manually; the contribution of \pkg{smoothbp} is to make it turnkey.

\end{document}